# Traffic noise assessment in urban Bulgaria using explainable machine learning

Marco Helbich[1,2,3,#,$], Julian Hagenauer[1,2,3,$], Angel Burov[2,3,4], Angel M. Dzhambov[2,3]

[1] Department of Human Geography and Spatial Planning, Faculty of Geosciences, Utrecht University, Utrecht, The Netherlands

[2] Health and Quality of Life in a Green and Sustainable Environment Research Group, Strategic Research and Innovation Program for the Development of MU – Plovdiv, Medical University of Plovdiv, Plovdiv, Bulgaria

[3] Environmental Health Division, Research Institute at Medical University of Plovdiv, Medical University of Plovdiv, Plovdiv, Bulgaria

[4] Department of Urban Planning, Faculty of Architecture, University of Architecture, Civil Engineering and Geodesy, Sofia, Bulgaria

[#] Corresponding author, Princetonlaan 8a, 3584 CB Utrecht, The Netherlands, Tel.: (0031) 30 253 2017, Email: m.helbich@uu.nl

[$] Both authors contributed equally

## Abstract

Fine-grained noise maps are vital for epidemiological studies on traffic noise. However, detailed information on traffic noise is often limited, especially in Eastern Europe. Rigid linear noise land-use regressions are typically employed to estimate noise levels; however, machine learning likely offers more accurate noise predictions. We innovated by comparing the predictive accuracies of supervised machine learning models to estimate traffic noise levels across the five largest Bulgarian cities. In situ A-weighted equivalent continuous sound levels were obtained from 232 fixed-site monitors across these cities. We included transport- and land-use-related predictors using 50-1,000 m buffers. Extreme gradient boosting (XGB) had the highest ten-fold cross-validated fit ($R^2$=0.680) and the lowest root mean square error (RMSE=4.739), insignificantly besting the random forest-based model ($R^2$=0.667, RMSE=4.895). Support vector regression ($R^2$=0.633, RMSE=5.358), elastic net ($R^2$=0.568, RMSE=5.625), and linear regression ($R^2$=0.548, RMSE=5.569) performed significantly worse. Shapley values for the XGB showed that the length of major roads within 100 m buffers, footways within 50 m buffers, residential roads within 50 m buffers, and the number of buildings within 50 m buffers were important non-linear predictors. Our spatially resolved noise maps revealed striking geographic noise variations and that, on average, 96.8% of the urban population experiences harmful noise levels.

# 1 Introduction

Transport noise is a primary source of noise pollution (Mann and Singh, 2022) and an environmental stressor detrimental to human health (Welch et al., 2023; World Health Organization, 2018). Meta-analyses have substantiated that excessive traffic noise exposure puts people at risk for non-auditory adverse health outcomes, including elevated annoyance (Guski et al., 2017), blood pressure (Chen et al., 2023), mental illness (Lan et al., 2020), and sleep disturbance (Smith et al., 2022). Furthermore, urban noise is unevenly distributed (European Environment Agency, 2019), with some areas experiencing disproportionate burdens (Hayward and Helbich, 2024; Peris and Arguelles, 2023). Approximately 20% of Europeans, predominantly in urban areas, are exposed to unhealthy noise concentrations of more than 55 decibels (dB) during the day-evening-night period (European Environment Agency, 2021). In Bulgaria, the proportion is substantially worse, with 76% of urban inhabitants exposed to harmful road noise (European Environment Agency, 2024).

Monitoring traffic noise across cities comprehensively is often infeasible because measurements are time-consuming, expensive, and labor-intensive (Mann and Singh, 2022). However, the European Environmental Noise Directive 2002/49/EC obligates countries to conduct strategic noise assessments (European Environment Agency, 2019; Murphy and King, 2010). For Bulgarian cities with more than 100,000 inhabitants, these traffic noise maps are delivered only in five dB(A) isophones rather than continuously and have been rated as low quality (Khomenko et al., 2022). Thus, such noise maps are only conditionally suitable for epidemiological studies as they can lead to exposure misclassification due to limited exposure contrasts and statistical power. Additionally, as strategic noise assessments target noise levels in high-exposure areas, low-traffic areas may be underrepresented (Dzhambov et al., 2023).

Consequently, the accurate and continuous mapping of noise exposure is an area of ongoing research using various approaches (Khan et al., 2018; Meller et al., 2023). Acoustic models that rely on the physics of noise propagation are the gold standard (Chang et al., 2012; Murphy and King, 2010). However, their application is often limited to selected sites or street segments due to input parameter sensitivity (e.g., façade information) (Chang et al., 2019; Xie et al., 2011). Geostatistical interpolation-based models (e.g., kriging) using measurements solely from monitoring sites also face challenges as they over-smooth small-scale noise contrasts (Aumond et al., 2018; Harman et al., 2016).

To spatialize measurements retrieved from stationary monitoring sites, land-use regression (LUR) models (Briggs et al., 1997; Hoek et al., 2008; Ma et al., 2024) are gaining popularity for noise mapping. LUR models capture small-scale noise variations on the principle that noise levels are a function of environmental characteristics (e.g., traffic or land-use-related) occurring within the ambit of monitoring sites. The fitted LUR is subsequently applied to predict spatially resolved noise concentrations at unsampled locations. The few available LUR-based studies predominantly use linear regression models (LM) (Aguilera et al., 2015; Chang et al., 2019; Gharehchahi et al., 2024; Harouvi et al., 2018; Raess et al., 2021; Staab et al., 2022; Xie et al., 2011; Xu et al., 2022). Despite LMs performing reasonably well, they rely on strong statistical assumptions (e.g., linearities) and are prone to overfitting when there are many predictors and few monitoring sites available, possibly inflating the accuracy of the noise mapping (Basagaña et al., 2012; Wang et al., 2012).

Unlike LMs, supervised machine learning (ML) models are more flexible (Zhu et al., 2023) and may supersede traditional LURs in their predictive ability (Fallah-Shorshani et al., 2022; Liu et al., 2020; Yin et al., 2020). ML models can efficiently handle high-dimensional data, approximate possible non-linear associations, and incorporate variable interactions that are not explicitly defined. They are also less constrained by restrictive underlying assumptions (Zhu et al., 2023). Advancements in ML resulted in many algorithms, making choosing a well-performing model difficult (Fernández-Delgado et al., 2014; Sekeroglu et al., 2022). Few studies, and none in Europe, have used ML for noise mapping (Fallah-Shorshani et al., 2022; Liu et al., 2020; Yin et al., 2020). The results of these studies have suggested that ML often, though not always, outperforms traditional LURs. For example, a Canadian study showed that random forest (RF) models outperformed LMs (Liu et al., 2020).

State-of-the-art ML algorithms applied to noise mapping focus on prediction accuracy, with little attention paid to the interpretability of such 'black-box' models. To explicate ML models' noise predictions, model-agnostic explainable ML approaches, including SHapley Additive exPlanations (SHAP), appear promising (Belle and Papantonis, 2021; Lundberg et al., 2020). As a unified framework for model interpretability, SHAP values explain individual predictions and overall model behavior. SHAP can, for example, uncover non-linearities between noise levels and environmental predictors and determine the importance of specific observations and predictors, all critical aspects to understanding complex ML-based noise models (Lundberg and Lee, 2017; Štrumbelj and Kononenko, 2014). To overcome the constraints of existing noise modeling approaches, we aimed 1) to develop the first-of-its-kind high-resolution traffic noise maps based on fixed-site monitoring data in urban Bulgaria; 2) to systematically compare and explain the predictive accuracies of ML-based LURs to estimate traffic noise; and 3) to assess the proportion of people exposed to unhealthy traffic noise exceeding 55 dB(A).

## 2 Materials and methods

### 2.1 Study area

The study was conducted in Bulgaria, the fifth most noise-polluted European country by the percentage of the urban population exposed to unhealthy traffic noise (European Environment Agency, 2021). We included the administrative areas of the five largest cities, including Sofia (≈1,249,000 people), Plovdiv (≈347,000 people), Varna (≈333,000 people), Burgas (≈203,000 people), and Ruse (≈145,000 people). Supplementary Figure S1 shows the locations of the cities, each having a specific mix of land-use comprising residential, commercial, and industrial sites as well as different road types, from local roads to major arterial roads. Sofia, Varna, and Burgas have airports situated in their vicinities. Plovdiv, by contrast, has an airport located in the surroundings, likely having minimal impact on its residents. All the cities possess railway networks, and Sophia also has a public tram system.

### 2.2 Noise measurement campaign

Local public health offices (i.e., Regional Health Inspectorates) performed legally mandated noise measurements at similar locations. We obtained measurement data from 232 fixed monitoring sites (75 in Sofia, 45 in Plovdiv, 45 in Varna, 37 in Burgas, and 30 in Ruse). Monitoring sites were geocoded (2018) using global positioning system (GPS) devices to enable precise repeated measurements. The fixed measurement stations were adjacent to traffic lanes, industrial sites, and residential and recreational areas. To obtain noise measurements across differing land-use types, 40% of the monitoring sites were next to traffic lanes, 30% were at industrial sites, and 30% were in noise-limited areas. The measurement protocol specified a distance of approximately 7.5 m between each monitoring site and the adjacent roadway. Supplementary Figure S2 displays the geographic distribution of the sites.

### 2.3 Noise measurements

Field workers from Regional Health Inspectorates took measurements with calibrated sound level meters three times a day (at least two measurements were during peak traffic hours) over two daytime periods between 07.00 and 19.00 h following ISO 1996-2 protocol norms. Each measurement was taken approximately 1.5 m above the ground for 15-20 min. Multiple brands of sound level meters were in use, with their types varying across cities and over time. However, Brüel & Kjær type 2238 sound meters with 4188 microphones and 4231 sound calibrators exemplified commonly used systems.

Measurements were available in equivalent continuous sound levels (A-weighted, in dB) ($L_{Aeq}$). For each sampling point, we had sound level data for each year from 2018 to 2022. Due to the high correlation and stability of the measured sound levels over time, we used the average values from 2018 to 2022 as the dependent variable in our LUR models (Das et al., 2019). Since our measurement campaign included COVID-19 lockdowns, we also assessed the temporal stability of our annual noise measurements. The yearly means and standard deviations (SD) revealed no significant fluctuations (2018: 63.17 [SD ± 8.34]; 2019: 63.19 [SD ± 8.47]; 2020: 62.56 [SD ± 8.54]; 2021: 62.90 [SD ± 8.49]; 2022: 62.56 [SD ± 8.54]).

### 2.4 Candidate predictor variables

The selection of covariates was informed by the literature (Aguilera et al., 2015; Gharehchahi et al., 2024; Ragettli et al., 2016) but constrained by data availability. We primarily considered covariates related to road or land-use features. Table 1 shows the candidate predictors. Road-based predictors represented the primary noise sources with magnitudes depending on their expected traffic capacity. Data was downloaded in October 2023 on different road categories' centerlines from OpenStreetMap (OSM) (Arsanjani et al., 2015). We included Euclidean distances between measurement locations (or the cell centroids) and the closest roads (e.g., motorways, primary roads, or footways). Additionally, we centered circular buffers ranging from 50 m to 1,000 m on the measurement locations (or the cell centroids for the predictions) to assess the total lengths of different road types within buffer sizes. We believed that smaller buffers affected sound propagation more directly than larger ones that incorporated sparsely distributed road types (e.g., motorways).

The most recent 2018 pan-European Urban Atlas provided land-use features (Copernicus, 2024). The nomenclature includes 17 urban classes with a minimum mapping unit of 0.25 ha. We considered Euclidean distances and the presence of different land-use types, including green urban areas (i.e., 14100), other green spaces (i.e., 23000: pastures, 22000: permanent crops, and 31000: forests), and urban fabric (i.e., 11100: continuous urban fabric, 12100: industrial, commercial, public, military and private units). We also included distances to airports (i.e., 12400) and railroads (i.e., 12230), which potentially can contribute to noise emissions. Since noise levels increase with urban density, we also included the number of buildings (i.e., their centroids) obtained from the cadaster (2023) from the local mapping agencies within our buffers and the degree of artificially sealed areas within buffers. The latter was acquired from the European imperviousness layer (10 m resolution) for 2018 and measures the degree of sealed areas ranging from zero to one, with higher values referring to increased imperviousness density. Finally, to incorporate spatial dependency between the observations, we also included the locational coordinates for each site.

Table 1: Candidate predictors used in the LUR development for road traffic noise.

| Variable description | Abbreviations | Source |
| --- | --- | --- |
| Geocoded locations of the measurement sites (XY coordinates, EPSG: 7801) | X, Y | Authors |
| Euclidean distance to the nearest road of any class (in m) | DARoad | OSM |
| Euclidean distance to the nearest motorway, primary, or secondary road (in m) | DMRoad | OSM |
| Euclidean distance to the nearest motorway (in m) | DMWay | OSM |
| Euclidean distance to the nearest primary road (in m) | DPRoad | OSM |
| Euclidean distance to the nearest secondary road (in m) | DSRoad | OSM |
| Euclidean distance to the nearest tertiary road (in m) | DTRoad | OSM |
| Euclidean distance to the nearest residential road (in m) | DRRoad | OSM |
| Euclidean distance to the nearest footway (in m) | DFWay | OSM |
| Length (in m) of all roads within 50 m, 100 m, 200 m buffers | LARoad | OSM |
| Length (in m) of motorway, primary or secondary roads within 50 m, 100 m, 200 m buffers | LMRoad | OSM |
| Length (in m) of primary roads within 50 m, 100 m, 200 m, 300 m, 500 m, 1,000 m buffers | LPRoad | OSM |
| Length (in m) of motorway within 50 m, 100 m, 200 m, 300 m, 500 m, 1,000 m buffers | LMWay | OSM |
| Length (in m) of secondary roads within 50 m, 100 m, 200 m, 300 m, 500 m, 1,000 m buffers | LSRoad | OSM |
| Length (in m) of tertiary roads within 50 m, 100 m, 200 m, 300 m, 500 m, 1,000 m buffers | LTRoad | OSM |
| Length (in m) of residential roads within 50 m, 100 m, 200 m, 300 m, 500 m, 1,000 m buffers | LRRoad | OSM |
| Length (in m) of footways within 50 m, 100 m, 200 m, 300 m, 500 m, 1,000 m buffers | LFWay | OSM |
| Euclidean distance to the nearest airport (in m) | DAir | UA |
| Euclidean distance to the nearest railway (in m) | DRail | UA |
| Euclidean distance to the nearest green space (in m) | DGreen | UA |
| Euclidean distance to the nearest green space of other types (pastures, crops, or forests) (in m) | DOGreen | UA |
| Euclidean distance to the nearest continuous urban fabric (in m) | DUrban | UA |
| Euclidean distance to the nearest industrial, commercial, public, military, and private units (in m) | DOLU | UA |
| Mean imperviousness density within 50 m, 100 m, 200 m, 300 m, 500 m, 1,000 m buffers | Imp | UA |
| Number of building centroids within 50 m, 100 m, 200 m, 300 m, 500 m, 1,000 m buffers | Build | Cadastre |

Note that numbers reported after the abbreviation refer to the buffer size. OSM = OpenStreetMap, UA = Urban Atlas. For predictors that include motorways, the buffer size was restricted to the immediate surroundings being more directly affected by sound propagation.

### 2.5 Modeling approaches

*2.5.1 Considered algorithms*

We fitted a basic linear regression (LM) as our benchmark model (Aguilera et al., 2015; Gharehchahi et al., 2024; Ragettli et al., 2016). Variance inflation factors assessed predictor collinearity stepwise, with values >10 deemed problematic (Craney and Surles, 2002), and thereafter, we applied forward stepwise regression. Guided by comparative ML studies (Fernández-Delgado et al., 2014; Hagenauer et al., 2019; Sekeroglu et al., 2022), we also developed noise models based on four ML algorithms and all the predictors (exclusive the length of secondary roads within 50 m and 200 m buffers which were perfectly correlated with the other predictors): First, the elastic net (ENet) integrated the properties of lasso and ridge regression by permitting variable selection through coefficient shrinkage and handled predictor collinearity (Zou and Hastie, 2005). Second, the support vector regression (SVR) (Smola and Schölkopf, 2004) with a radial basis kernel used a hyperplane that fit best within a predefined margin around the data points transformed into a high-dimensional space where a linear approximation was feasible. Third, the random forest (RF) used an ensemble of regression trees (Breiman, 2001; Wright and Ziegler, 2017). For the training of each tree, a bootstrap sample of the data was selected. At each tree's splitting node, a random sample of features was selected. After training the unpruned trees, the predictions were averaged across the trees. Fourth, we applied extreme gradient boosting (XGB), which considers an ensemble of sequentially trained weak learners, typically regression trees (Chen and Guestrin, 2016). Iteratively, each tree was trained on the residuals of the previous one, gradually enhancing the prediction accuracy. We combined the predictions from all the trees to obtain the final predictions. Since ML models tend to perform better with many observations, we pooled the data rather than conducting city-specific analyses.

*2.5.2 Model tuning, validation, and selection*

A robust model that does not overfit is a prerequisite to obtaining accurate predictions. Except for basic regression, ML models have hyperparameters that require tuning before assessing the prediction accuracy of the trained model (Kohavi, 1995). We applied ten-fold cross-validation (CV) to determine the best hyperparameters of each model nested within four times repeated ten-fold CV to assess the predictive performance. Repeating the CV reduces sampling variability and ensures that the predictive performance is less dependent on a particular random split. Supplementary Table S1 summarizes the tested hyperparameters, and Table S2 the selected ones. As the predictors were skewed, we applied the Yeo-Johnson transformation to obtain more symmetrically distributed predictors (Yeo and Johnson, 2000).

We reported three performance metrics between the observed and predicted values: the root mean square error (RMSE), the mean absolute error (MAE), and the coefficient of determination ($R^2$). To test whether the models' predictions differed statistically from one another by applying two-tailed Wilcoxon rank-sum tests with the Benjamini–Hochberg correction for multiple testing. If $p<0.05$, the models significantly differed.

When fitting aspatial models using geospatial data, it is critical for inference that the residuals are spatially independent. We used Moran's *I* statistic to examine residual spatial independence based on an inverse distance weight matrix specification. We tested significance over 999 Monte Carlo simulations. The analyses were done in R (R Core Team, 2024) with the *leaps*, *elasticnet, kernlab, ranger*, and *xgboost* package. The *caret* package provided an interface for the algorithms' parameter tuning and validation(Kuhn et al., 2013). For the reproducibility of our analysis, we provide the R code as supplementary files.

*2.5.3 Model interpretation*

To overcome the hurdle of model interpretability, we opted for an in-depth understanding of the best-performing model. As a post hoc approach, we employed SHapley Additive exPlanations (SHAP), as implemented in the R package *imp* (Molnar et al., 2018). SHAP is a unified approach to explaining ML predictions by generating locally additive feature attribution (Lundberg et al., 2020). Using a trained model, SHAP decomposes predictions into observation-specific explanations by capturing the average attribution for each observation to the model's prediction in every possible order (Lundberg and Lee, 2017).

We used SHAP values in four ways (Molnar et al., 2018): First, to capture the global predictor importance, we determined the mean absolute SHAP values for each one. Additionally, we assessed each predictor's mean absolute SHAP values for each city to explore possible between-city differences in terms of predictor importance. The higher the absolute SHAP value, the more critical the predictor. Second, we generated beeswarm plots, allowing further deconstruction of how each observation affected the prediction for each city. Higher SHAP values referred to greater effects (positive or negative), with negative SHAP values exerting negative effects on outcomes and vice versa. Third, we used SHAP dependence plots to assess the shape of the predictors-outcome associations. The SHAP analyses were limited to the eight most important predictors.

*2.5.4 Traffic noise predictions*

We mapped the traffic noise estimates of the best-performing model. We superimposed a 50 m grid on each city to obtain predicted noise levels in dB(A) and used Leaflet to create an interactive noise map to facilitate data exploration. We also estimated the proportions of the population exposed to traffic noise levels ranging from 40 to 70 dB(A) in 5 dB(A) isophones. National census data from 2021 provided the population estimates.

**3 Results**

*3.1 Noise measurements*

We used in-situ noise measurements from 232 sites. The mean noise level across the measurements taken in the cities was 63.0 dB(A) (SD ± 8.2). The average noise levels taken in a city ranged from a minimum of 39.7 to a maximum of 75.3 db(A), representing a 35.6 dB(A) difference. Mean noise distributions varied across and within the cities (Figure 1) despite only moderate geographic variations. Observed mean noise levels were highest in Plovdiv (68.36 ± 2.97), followed by Ruse (65.45 ± 5.71), Burgas (64.81 ± 9.77), Sofia (61.53 ± 7.98), and Varna (57.06 ± 7.95). Across all cities, inner-city areas experienced higher noise levels than suburban areas.

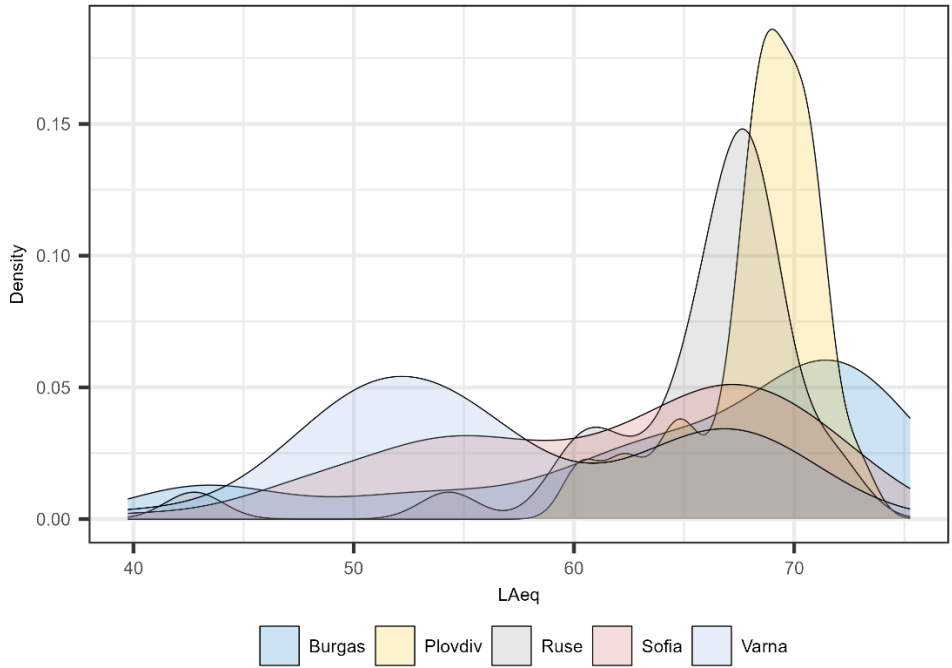

Figure 1: Equivalent continuous noise distribution for each city ($L_{Aeq}$).

### 3.2 Cross-validated predictive accuracy

Figure 2 and Supplementary Table S3 summarize the fit statistics for the five competing models based on the four times repeated ten-fold CV after hyperparameter tuning. All models achieved good median cross-validated fits, regardless of the measure. However, the XGB model exhibited the best fit, followed by the RF model. The XGB model achieved the highest $R^2$ of 0.680 and the lowest RMSE of 4.739 dB(A), followed by the RF ($R^2$=0.665; RMSE=4.895), the SVR ($R^2$=0.633; RMSE=5.358), and the ENet ($R^2$=0.568; RMSE=5.625). The difference between the XGB and RF models was statistically insignificant based on the Wilcoxon test ($p$>0.05), but XGB significantly outperformed the remaining models (all $p$<0.05). The traditional LM with forward predictor selection (based on the Euclidean distance to the nearest motorway, primary, or secondary road, the nearest railway, imperviousness density within 200 m buffers, tertiary roads within 50 m, and the Y coordinate) performed the least well among the tested models, with an $R^2$ of 0.548 (RMSE=5.569). The MAEs mimicked the results of the RMSEs.

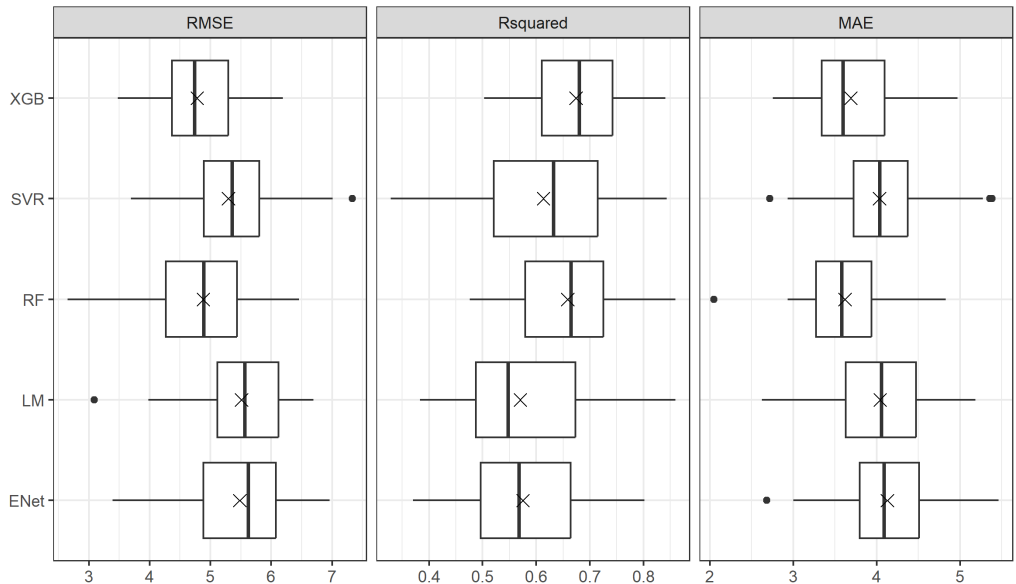

Figure 2: Fit statistics across the LUR models based on four times repeated ten-fold CV with hyperparameters selected based on nested ten-fold CV. The crosses represent the means, and the dots represent the outliers.

Supplementary Figure S3 depicts RMSE and MAE values for each city based on the pooled data. Regardless of the model examined, the boxplots indicated that the prediction errors varied by city. For example, the XGB model had the lowest RMSE and MAE values in Plovdiv and Ruse, while the prediction errors were greatest in Varna and Sofia. Figure 3 serves as an in-sample model diagnostic of how well the predicted noise levels matched the measured ones. We did not observe any systematic over- or underpredictions for the XGB and the RF model. However, the spread between the predicted and measured noise levels increased with decreasing $R^2$ (see LM). XGB's residual Moran's *I* was -0.039 ($p$=0.186). None of the other models faced spatially autocorrelated residuals (Supplementary Table S4).

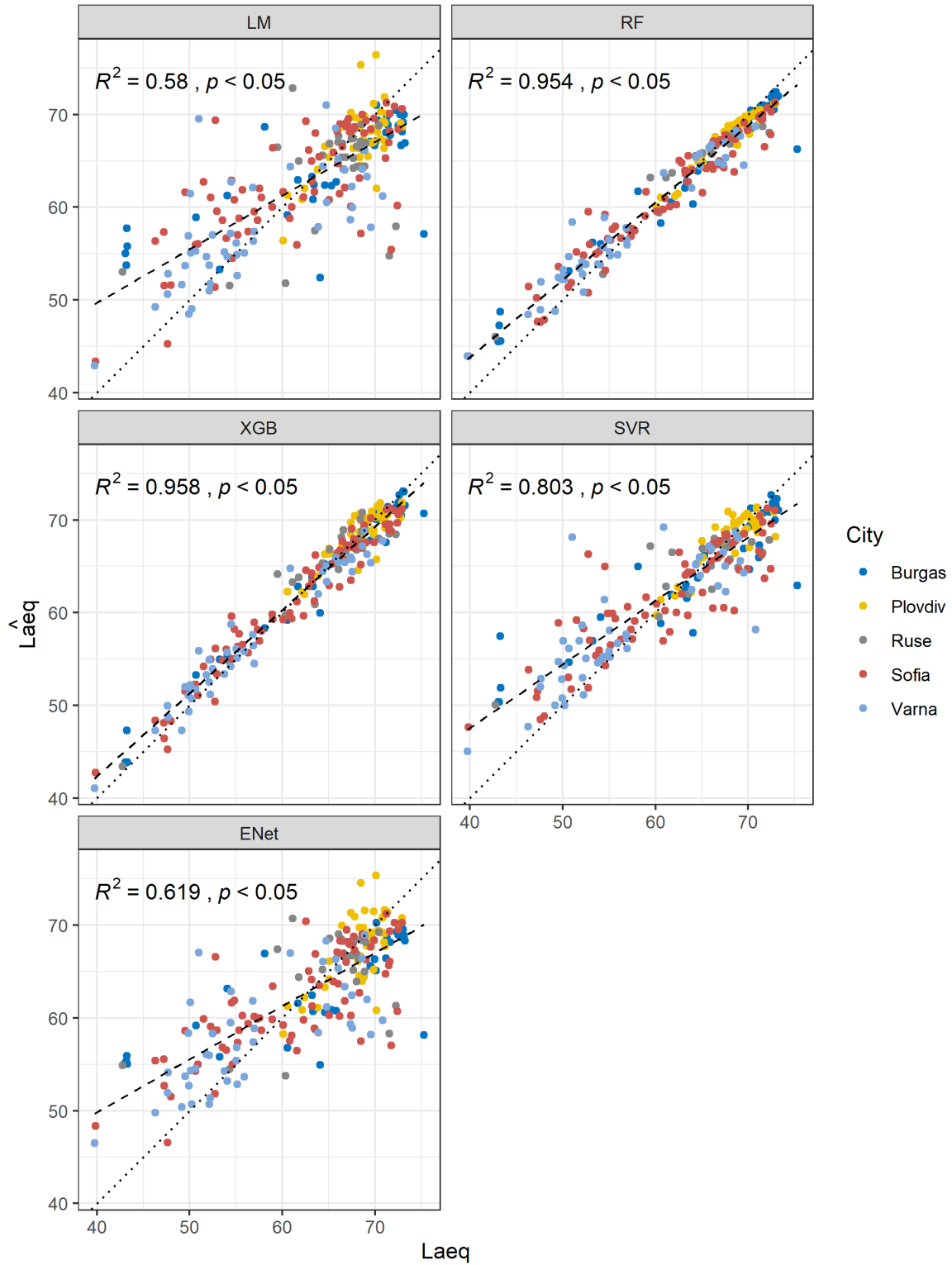

Figure 3: In-sample comparison of predicted and measured noise levels at the measurement sites. The colors of the data points identify the cities where the measurement sites are located. The strengths of the associations are measured through the correlation coefficient squared ($R^2$). The dashed line represents the regression line. The dotted line represents the 1:1 line.

*3.3 Model explanation*

Figure 4 summarizes the most important predictors of the mean absolute SHAP values from the best-performing XGB model. Across cities, the length of major roads within 100 m buffers was consistently the most important predictor, followed by the length of footways within 50 m buffers, and the length of residential roads within 50 m buffers. We noted that the magnitude of the variables and, hence, their importance differed substantially across the cities (Figure 4). For example, the Euclidean distance to the nearest continuous urban fabric in Plovdiv was approximately twice the magnitude found in Varna and Sofia.

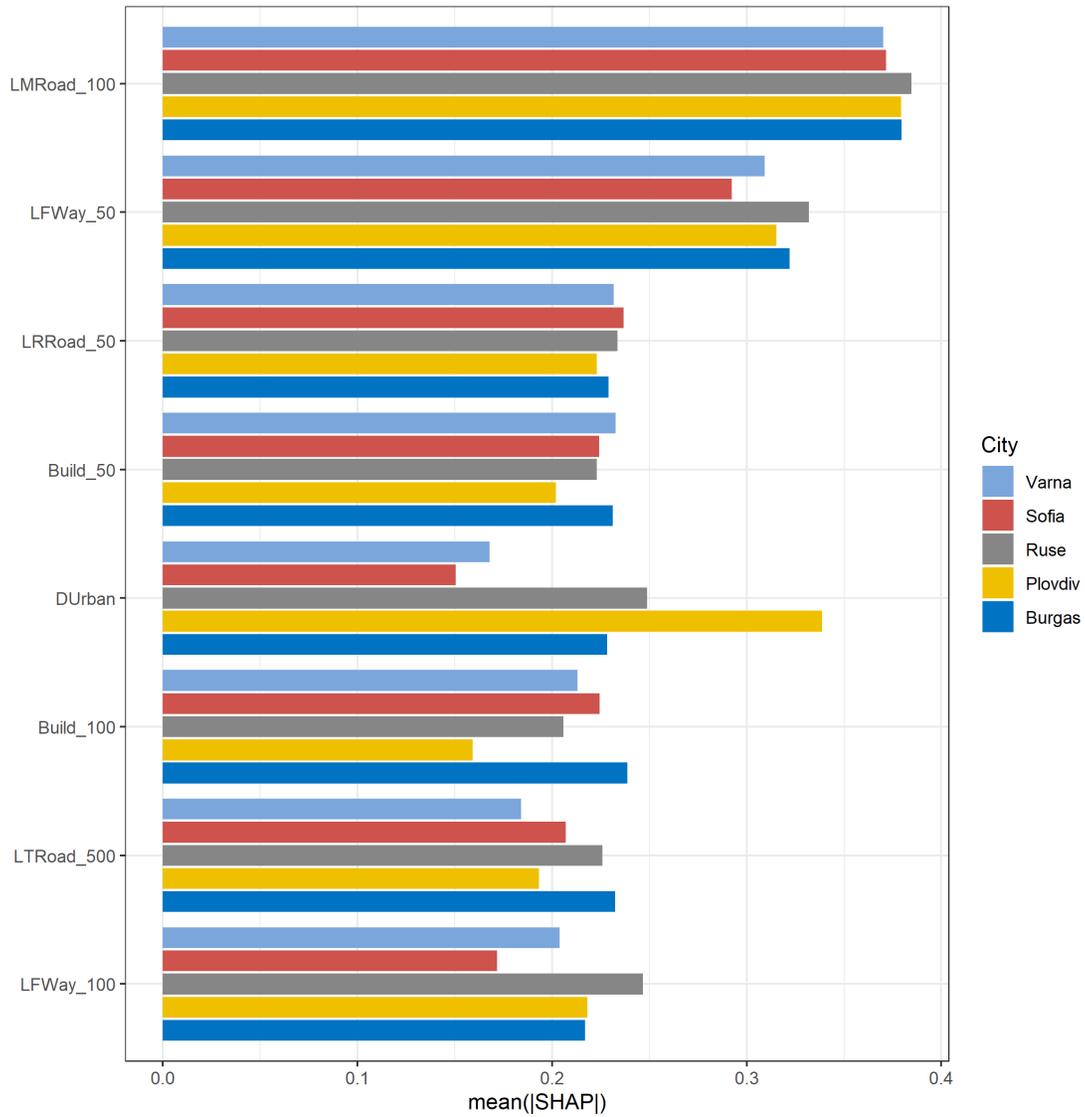

Figure 4: Predictor importance based on the SHAP values by city for XGB. While the SHAP values are colored by city, the underlying model is based on the pooled data. The plot is restricted to the eight most important predictors. The predictor relevance decreases from the top to the bottom.

Figure 5 explains how each observation affects the XGB model predictions. Figures 5 (and Supplementary Figure S4) display dichotomous distributions of the SHAP values. To capture geographic differences, we considered each city separately. When the variable values exceed 0, the SHAP values, depending on the variable, are either substantially higher or lower than 0 and maintain this trend throughout the remaining value range. Regardless of the city, it is noteworthy that the predictors featured positive and negative observations of noise effects, although the specific patterns appeared to be city-dependent. For example, the length of major roads within 100 m buffers related to observations that primarily had positive effects on noise in Burgas and Plovdiv, while in Varna negative SHAP values dominate. To ascertain the shape of the associations of the eight most important noise predictors, Supplementary Figure S4 shows SHAP dependence plots. The SHAP values of the length of major roads within 100 m buffers showed a positive association up to 70 m before leveling off. The negative association between noise levels and the length of footways within 50 m buffers drops off within a short distance.

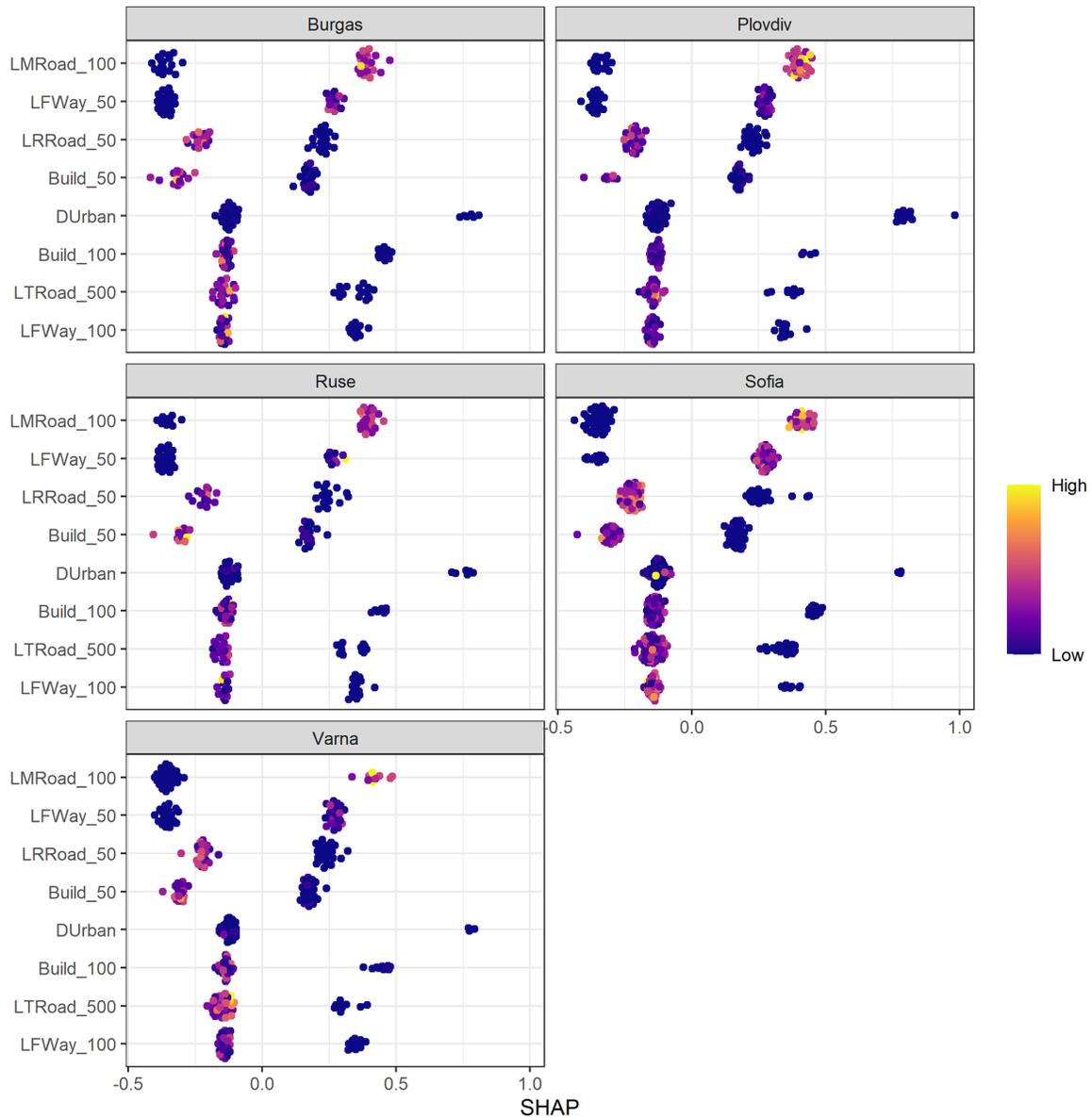

Figure 5: Distribution of SHAP values by feature and city for XGB. While the SHAP values are stratified by city, the underlying model is based on pooled data. The color gradient indicates the normalized predictor values. Predictors are in descending order of importance. Positive SHAP values indicate a positive effect on noise, and negative ones have a negative effect. The higher the SHAP value, the stronger the effect.

### 3.4 Traffic noise mapping

Figure 6 depicts the city-specific traffic noise predictions on a 50 m grid using the XGB model. For comparative purposes, Supplementary Figures S5-S9 also show the noise surfaces based on the less-well-performing algorithms. The noise estimates using XGB showed that the model captures noise level variations effectively, particularly close to major roads. Traffic noise levels tended to be higher in inner cities and areas with manufacturing industries, commercial activities, and industry versus more rural areas.

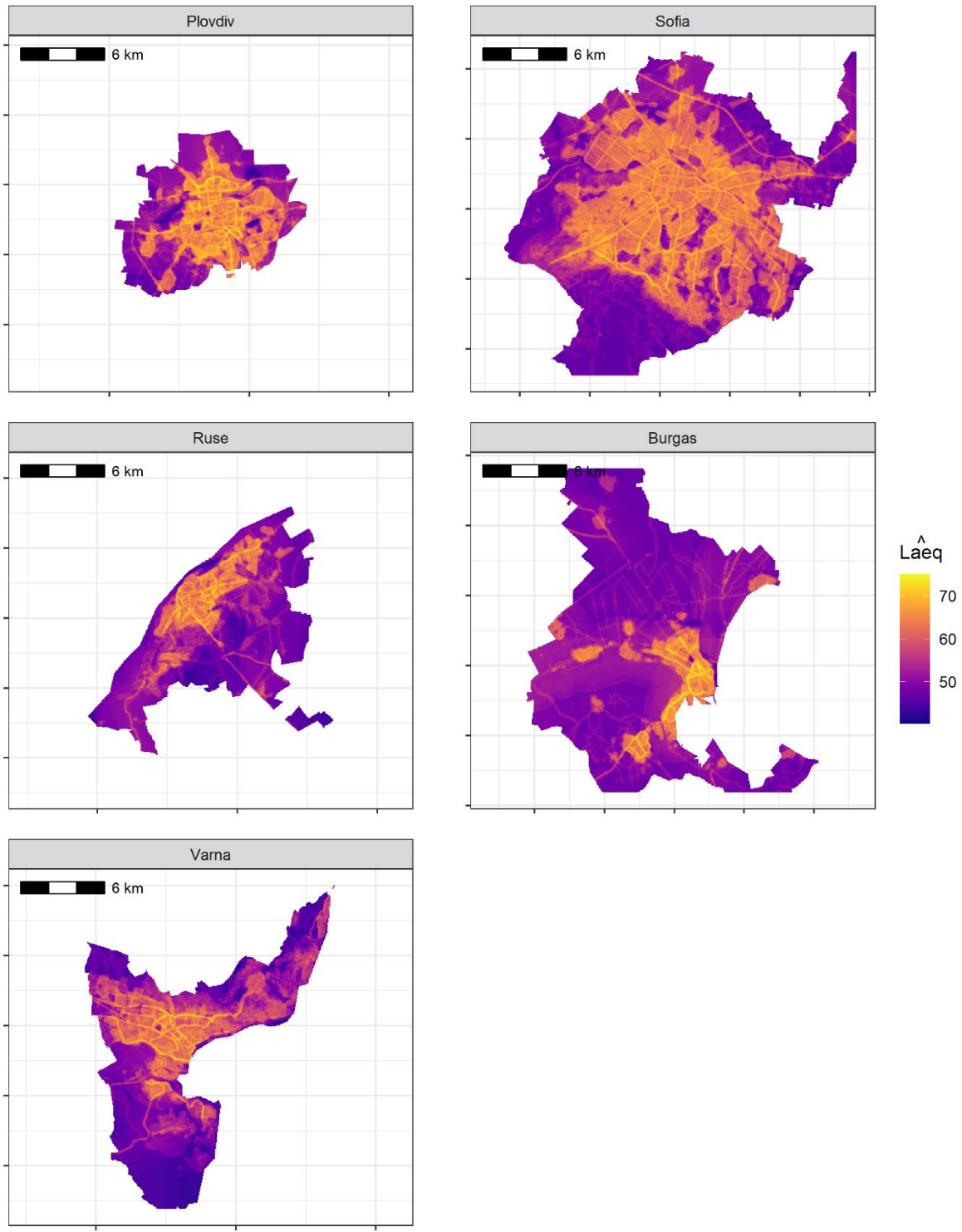

Figure 6: Predicted traffic noise levels in dB(A) based on the XGB model. Interactive maps are provided in a Supplementary File.

Table 2 summarizes the proportion and number of the population exposed to different traffic noise levels. On average, approximately 96.8% of the population experienced exposure to long-term noise levels of >55 dB(A), but the percentage differs across cities (e.g., 83.8% in Varna and 99.7% in Plovdiv). Supplementary Figure S10 depicts the population's noise level exposure estimations for various models.

Table 2: Estimated proportion and number of the population exposed to different traffic noise levels in total and stratified by city. The estimates rely on the XGB noise predictions.

| dB(A) | Population affected (%/total) | Population affected per city (%/total) | | | | |
|---|---|---|---|---|---|---|
| | | Burgas | Plovdiv | Ruse | Sofia | Varna |
| >40 | 100.0% | 100.0% | 100.0% | 100.0% | 100.0% | 100.0% |
| | 2,112,821 | 188,889 | 321,962 | 124,102 | 1,171,247 | 306,621 |
| >45 | 100.0% | 100.0% | 100.0% | 100.0% | 100.0% | 99.9% |
| | 2,112,577 | 188,886 | 321,959 | 124,102 | 1,171,247 | 306,383 |
| >50 | 99.4% | 99.7% | 100.0% | 99.8% | 99.8% | 96.8% |
| | 2,100,263 | 188,375 | 321,915 | 123,910 | 1,169,342 | 296,721 |
| >55 | 96.8% | 99.4% | 99.7% | 97.9% | 98.9% | 83.8% |
| | 2,045,116 | 187,707 | 321,077 | 121,555 | 1,157,854 | 256,923 |
| >60 | 81.8% | 93.6% | 97.3% | 90.6% | 83.7% | 47.5% |
| | 1,728,041 | 176,859 | 313,260 | 112,448 | 979,749 | 145,725 |
| 65 | 24.9% | 48.0% | 58.3% | 25.9% | 16.7% | 6.6% |
| | 525,776 | 90,654 | 187,600 | 32,085 | 195,191 | 20,246 |
| >70 | 1.6% | 6.8% | 5.5% | 0.3% | 0.1% | 0.1% |
| | 32,859 | 12,778 | 17,734 | 355 | 1,589 | 403 |

4 Discussion

Long-term exposure to traffic noise has been associated with multiple adverse health effects (Chen et al., 2023). Such evidence relies on fine-grained noise exposure surfaces, which are not readily available in all places, including Bulgaria. Our study overcame the rigidities associated with linear noise LURs (Aguilera et al., 2015; Chang et al., 2019; Gharehchahi et al., 2024; Harouvi et al., 2018; Raess et al., 2021; Xie et al., 2011) by developing ML-based models to estimate traffic noise.

*4.1 Principal findings and available evidence*

Our results showed that ensemble-based tree models demonstrated strong performance, with standout performance from the XGB model, accounting for 68.0% of traffic noise variability. Excepting the RF model, XGB significantly outperformed all competing models. Unlike the LM, a model restricted to additive effects, tree-based models are well-suited for noise predictions because they effectively handle non-linearities and interactions across the predictors while minimizing model assumptions (Breiman, 2001; Chen and Guestrin, 2016). Our benchmark results can assist in model preselection but are not universally applicable to all LUR models, as performance is also affected by the type of data available and may be context-dependent.

Our results were generally consistent with the few available studies similar to ours and aligned with widely recognized benchmarks for tree boosting (Chen and Guestrin, 2016). In a mobile mapping study in Long Beach, CA, XGB outperformed acoustic models (e.g., CadnaA) in predicting noise levels (Fallah-Shorshani et al., 2022). Despite adjusting for traffic volume, a key noise emitter (Yin et al., 2020), the $R^2$ on a 30% retained test set was comparable to our result (Fallah-Shorshani et al., 2022). However, a study of three mid-sized European cities found that LUR fits relying solely on geographic information system (GIS)-based predictors are only negligibly smaller than when including auxiliary predictors collected during measurement site visits (Aguilera et al., 2015). A study of Canadian cities that compared an RF with a LM resulted in a substantially lower leave-one-out cross-validated $R^2$ of 0.58 for the RF ($R^2_{LUR}$=0.47) than in our study and supported our findings that tree-based models perform well (Liu et al., 2020). Finally, in a study in Sao Paulo, Brazil, the fit of a LM was also lower than our fit (Raess et al., 2021). Although our models relied on measurement data, which are not universally accessible, experiments with an end-to-end deep learning model to approximate noise levels from Sentinel-2 satellite imagery demonstrated encouraging results (RMSE≈8.2 dB), but predictions in urban areas continue to pose difficulties (Eicher et al., 2022). Another deep learning-based noise study achieved an overall accuracy of 0.89 (Staab et al., 2023). Given the differences between study sites, model validation (ten-fold vs. leave-one-out CV), measurement duration, and measurement type (stationary vs. mobile), close comparisons across the available evidence must be made cautiously.

Tree-based models can be challenging to interpret. To overcome this, we made the first LUR-related attempt to use SHAP values to deconstruct the black-box nature of ensemble models. Though computationally intensive, SHAP values are attractive because they are model-independent, resulting in a consistent predictor ranking, in contrast to other feature attribution approaches (e.g., gain) (Lundberg et al., 2020; Lundberg and Lee, 2017). In our study, the SHAP values suggested that the length of major roads within 100 m buffers, footways within 50 m buffers, residential roads within 50 m buffers, and the number of buildings within 50 m buffers were among the most important predictors. Given the evidence from other LUR-based noise studies, there is no one-fits-all predictor set. However, some similar predictors also appeared among the top predictors elsewhere (Fallah-Shorshani et al., 2022). Unlike in two Canadian studies (Liu et al., 2020; Ragettli et al., 2016), our results did not show that the presence of green space, thought to attenuate traffic noise (Klingberg et al., 2017), is among the most critical variables. This discrepancy may be due to geographic differences in terms of urban morphology, road network structure, or various possible green space operationalizations. There were two additional differences between our study and its predecessors. First, we quantified available green space using land-use data rather than using the normalized difference vegetation index. Second, we calculated SHAP values and found that the predictors are somehow non-linearly related to traffic noise, a fact not always considered in other LUR models (Aguilera et al., 2015; Chang et al., 2019; Raess et al., 2021; Xie et al., 2011).

For our traffic noise surfaces, we observed significant differences across the models related to granularity, echoing prior findings (Fallah-Shorshani et al., 2022; Liu et al., 2020). Compared to the LM, the XGB model surface showed more variation, likely consequent to incorporating predictor interactions and non-linearities. Fine-grained noise maps such as ours are vital for promoting urban sustainability. For example, our timely analysis provides the foundation for noise-related health impact assessments (Khomenko et al., 2022) and identifying noise pollution hotspots that should be prioritized for targeted interventions (e.g., installation of sound barriers). Such strategies are critical to attaining the 2030 target for a 30% reduction (versus 2017) in Europeans chronically impacted by transport noise, as envisaged in the European zero-pollution action plan (European Environment Agency, 2022). However, despite the European noise guidelines recommending keeping traffic noise levels below 55 dB(A) to minimize detrimental health effects (World Health Organization, 2018), we found that, on average, 96.8% of the population in our study areas was exposed to long-term noise levels that exceeded this threshold, implying attendant increases in health risks.

### 4.2 Strengths and limitations

A strength of our study is the substantial number of noise measurement locations over an extended period, allowing the capture of daily and seasonal variations by standardized protocols. Our study is one of the few using ML to calibrate LUR for noise assessment. Rather than arbitrarily selecting a single ML model (Liu et al., 2020), we evaluated several algorithms with different assumptions. Relatedly, we minimized the risk of overfitting and prevented overly optimistic performance estimates by rigorously tuning the hyperparameters rather than relying on off-the-shelf settings, which could open the possibility of inflating model performance. Whereas other ML-based LUR studies concentrated solely on predictive modeling performance (Fallah-Shorshani et al., 2022), we utilized Shapley values to understand our model better. By explicitly incorporating our measurement sites' locational information, our models could account for spatial effects. Finally, we adhered to the open data principles and made our noise estimates public to facilitate their reuse in future health and health impact assessments.

Some limitations are inherent to our analyses, typical of LUR studies, including ours. While our noise metric only accounted for daytime ($L_{day}$), it can be adjusted to reflect the average sound level over a 24-hour period by adding a 5 dB penalty for evening hours (19.00 to 22.00 h) and a 10 dB penalty for nighttime hours (22.00 to 07.00 h) using an empirically derived conversion term (Brink et al., 2018). While we focused on four well-established ML algorithms within the exposure assessment field, we emphasize that other ones, some of which may perform better, are available (Fernández-Delgado et al., 2014; Hagenauer et al., 2019; Sekeroglu et al., 2022). Our noise model does not depend on the physics of acoustics, which likely restricts its transferability to other cities. Although there is no universally accepted predictor set, a limitation is the lack of explanatory variables suspected to be associated with noise, although our model achieved a good fit compared to others with traffic volume data (Fallah-Shorshani et al., 2022). As our primary source of predictors, we used crowd-sourced OpenStreetMap data, which is susceptible to data quality issues (Arsanjani et al., 2015). Thus, we cannot exclude the possibility of inaccuracies in that data. However, OpenStreetMap has proven to be a reliable data source of high positional accuracy (Helbich et al., 2012), often used in LUR studies (Raess et al., 2021; Staab et al., 2022). Our measurements and land-use data were temporally misaligned. We believe, however, that the consequence for our LUR models was negligible since land-use dynamics, especially in cities, tend to be stable over timespans as short as five years. Finally, our grid-based noise predictions may induce some inaccuracies due to small-scale spatial variability of noise levels (e.g., two street segments in close proximity may have different noise levels despite falling within the same cell).

5. Conclusion

We rigorously compared ML algorithms for estimating long-term traffic noise levels across the five largest Bulgarian cities, for which detailed noise estimates were unavailable. Our XGB model explained 68.0% of the noise variation, slightly superior to RF and significantly better than SVR, ENet, and LM. Given the XGB model's high flexibility and superior performance, in combination with the ability to use SHAP values for model insights, we recommended the XGB model for future noise mappings. Linking the noise estimates to population data, we found that 96.8% of the urban population was at risk of experiencing harmful traffic noise. Our spatially resolved noise surfaces are openly available and provide opportunities for reuse in epidemiological studies assessing noise effects on human health. Policymakers can access the maps in support of developing new mitigation strategies for combatting exposure to traffic noise.

Abbreviations

CV = cross-validation

$L_{den}$ = day-evening-night period

ENet = elastic net

XGB = extreme gradient boosting

GIS = geographic information system

GPS = global positioning system

LUR = land-use regression

ML = machine learning

MAE = mean absolute error

OSM = OpenStreetMap

RF = random forest

LM = linear regression

RMSE = root mean square error

SHAP = SHapley Additive exPlanations

SVR = support vector regression

UA = Urban Atlas

## Acknowledgments

We thank the Regional Health Inspectorates for sharing their noise measurements. We also thank the reviewers for the constructive comments that have enhanced the quality of the manuscript.

## CRediT authorship contribution statement

MH: Conceptualization, Methodology, Visualization, Writing - Original Draft; JH: Conceptualization, Methodology, Formal analysis, Visualization, Writing - Review & Editing; AB: Data Curation, Writing - Review & Editing; AMD: Conceptualization, Data Curation, Writing - Review & Editing, Project administration, Funding acquisition

## Funding

The research leading to this work was supported by the "Strategic research and innovation program for the development of Medical University – Plovdiv" No. BG-RRP-2.004-0007-C01, Establishment of a network of research higher schools, National plan for recovery and resilience, financed by the European Union – NextGenerationEU. The funder did not influence the study design, data collection and analysis, interpretation, or article drafting. All authors had data access. The first and last authors were responsible for submitting the article for publication.

## Conflict of interest

The authors have no conflicts of interest to declare.

## Declaration of generative AI in scientific writing

No generative AI and AI-assisted technologies were used during the preparation of this work.

## Data sharing

The code and data generated and analyzed during the current study are available through a repository (DOI: TO_BE_ADDED_AFTER_REVIEW) and can be requested from the authors.

## Appendix. Supplementary materials

Supplementary materials to this article can be found online at LINK_TO_BE_ADDED_HERE

## References

Aguilera, I., Foraster, M., Basagaña, X., Corradi, E., Deltell, A., Morelli, X., Phuleria, H.C., Ragettli, M.S., Rivera, M., Thomasson, A., others, 2015. Application of land use regression modelling to assess the spatial distribution of road traffic noise in three European cities. J Expo Sci Environ Epidemiol 25, 97–105.

Arsanjani, J.J., Mooney, P., Helbich, M., Zipf, A., 2015. An exploration of future patterns of the contributions to OpenStreetMap and development of a Contribution Index. Transactions in GIS 19, 896–914.

Aumond, P., Can, A., Mallet, V., De Coensel, B., Ribeiro, C., Botteldooren, D., Lavandier, C., 2018. Kriging-based spatial interpolation from measurements for sound level mapping in urban areas. J Acoust Soc Am 143, 2847–2857.

Basagaña, X., Rivera, M., Aguilera, I., Agis, D., Bouso, L., Elosua, R., Foraster, M., de Nazelle, A., Nieuwenhuijsen, M., Vila, J., others, 2012. Effect of the number of measurement sites on land use regression models in estimating local air pollution. Atmos Environ 54, 634–642.

Belle, V., Papantonis, I., 2021. Principles and practice of explainable machine learning. Front Big Data 4, 688969.

Breiman, L., 2001. Random forests. Mach Learn 45, 5–32.

Briggs, D.J., Collins, S., Elliott, P., Fischer, P., Kingham, S., Lebret, E., Pryl, K., Van Reeuwijk, H., Smallbone, K., Van Der Veen, A., 1997. Mapping urban air pollution using GIS: a regression-based approach. International Journal of Geographical Information Science 11, 699–718.

Brink, M., Schäffer, B., Pieren, R., Wunderli, J.M., 2018. Conversion between noise exposure indicators Leq24h, LDay, LEvening, LNight, Ldn and Lden: Principles and practical guidance. Int J Hyg Environ Health 221, 54–63.

Chang, T.-Y., Liang, C.-H., Wu, C.-F., Chang, L.-T., 2019. Application of land-use regression models to estimate sound pressure levels and frequency components of road traffic noise in Taichung, Taiwan. Environ Int 131, 104959.

Chang, T.-Y., Lin, H.-C., Yang, W.-T., Bao, B.-Y., Chan, C.-C., 2012. A modified Nordic prediction model of road traffic noise in a Taiwanese city with significant motorcycle traffic. Science of the Total Environment 432, 375–381.

Chen, T., Guestrin, C., 2016. Xgboost: A scalable tree boosting system, in: Proceedings of the 22nd Acm Sigkdd International Conference on Knowledge Discovery and Data Mining. pp. 785–794.

Chen, X., Liu, M., Zuo, L., Wu, X., Chen, M., Li, X., An, T., Chen, L., Xu, W., Peng, S., others, 2023. Environmental noise exposure and health outcomes: an umbrella review of systematic reviews and meta-analysis. Eur J Public Health 33, 725–731.

Copernicus, 2024. Urban Atlas 2018 [WWW Document]. https://land.copernicus.eu/en/products/urban-atlas.

Craney, T.A., Surles, J.G., 2002. Model-dependent variance inflation factor cutoff values. Qual Eng 14, 391–403.

Das, P., Talukdar, S., Ziaul, S.K., Das, S., Pal, S., 2019. Noise mapping and assessing vulnerability in meso level urban environment of Eastern India. Sustain Cities Soc 46, 101416.

Dzhambov, A.M., Dimitrova, V., Germanova, N., Burov, A., Brezov, D., Hlebarov, I., Dimitrova, R., 2023. Joint associations and pathways from greenspace, traffic-related air pollution, and noise to poor self-rated general health: A population-based study in Sofia, Bulgaria. Environ Res 231, 116087.

Eicher, L., Mommert, M., Borth, D., 2022. Traffic Noise Estimation from Satellite Imagery with Deep Learning, in: IGARSS 2022-2022 IEEE International Geoscience and Remote Sensing Symposium. pp. 5937–5940.

European Environment Agency, 2024. Noise data reported under Environmental Noise Directive [WWW Document].

European Environment Agency, 2022. Health impacts of exposure to noise from transport.

European Environment Agency, 2021. Exposure of Europe's population to environmental noise.

European Environment Agency, 2019. Environmental noise in Europe — 2020.

Fallah-Shorshani, M., Yin, X., McConnell, R., Fruin, S., Franklin, M., 2022. Estimating traffic noise over a large urban area: An evaluation of methods. Environ Int 170, 107583.

Fernández-Delgado, M., Cernadas, E., Barro, S., Amorim, D., 2014. Do we need hundreds of classifiers to solve real world classification problems? The journal of machine learning research 15, 3133–3181.

Gharehchahi, E., Hashemi, H., Yunesian, M., Samaei, M., Azhdarpoor, A., Oliaei, M., Hoseini, M., 2024. Geospatial Analysis for Environmental Noise Mapping: A Land Use Regression Approach in a Metropolitan City. Environ Res 119375.

Guski, R., Schreckenberg, D., Schuemer, R., 2017. WHO environmental noise guidelines for the European region: A systematic review on environmental noise and annoyance. Int J Environ Res Public Health 14, 1539.

Hagenauer, J., Omrani, H., Helbich, M., 2019. Assessing the performance of 38 machine learning models: the case of land consumption rates in Bavaria, Germany. International Journal of Geographical Information Science 33, 1399–1419.

Harman, B.I., Koseoglu, H., Yigit, C.O., 2016. Performance evaluation of IDW, Kriging and multiquadric interpolation methods in producing noise mapping: A case study at the city of Isparta, Turkey. Applied Acoustics 112, 147–157.

Harouvi, O., Ben-Elia, E., Factor, R., de Hoogh, K., Kloog, I., 2018. Noise estimation model development using high-resolution transportation and land use regression. J Expo Sci Environ Epidemiol 28, 559–567.

Hayward, M., Helbich, M., 2024. Environmental noise is positively associated with socioeconomically less privileged neighborhoods in the Netherlands. Environ Res 248, 118294.

Helbich, M., Amelunxen, C., Neis, P., Zipf, A., 2012. Comparative spatial analysis of positional accuracy of OpenStreetMap and proprietary geodata. Proceedings of GI_Forum 4, 24.

Hoek, G., Beelen, R., De Hoogh, K., Vienneau, D., Gulliver, J., Fischer, P., Briggs, D., 2008. A review of land-use regression models to assess spatial variation of outdoor air pollution. Atmos Environ 42, 7561–7578.

Khan, J., Ketzel, M., Kakosimos, K., Sørensen, M., Jensen, S.S., 2018. Road traffic air and noise pollution exposure assessment–A review of tools and techniques. Science of the total environment 634, 661–676.

Khomenko, S., Cirach, M., Barrera-Gómez, J., Pereira-Barboza, E., Iungman, T., Mueller, N., Foraster, M., Tonne, C., Thondoo, M., Jephcote, C., others, 2022. Impact of road traffic noise on annoyance and preventable mortality in European cities: A health impact assessment. Environ Int 162, 107160.

Klingberg, J., Broberg, M., Strandberg, B., Thorsson, P., Pleijel, H., 2017. Influence of urban vegetation on air pollution and noise exposure–a case study in Gothenburg, Sweden. Science of the Total Environment 599, 1728–1739.

Kohavi, R., 1995. A study of cross-validation and bootstrap for accuracy estimation and model selection, in: Ijcai. pp. 1137–1145.

Kuhn, M., Johnson, K., others, 2013. Applied predictive modeling. Springer.

Lan, Y., Roberts, H., Kwan, M.-P., Helbich, M., 2020. Transportation noise exposure and anxiety: A systematic review and meta-analysis. Environ Res 191, 110118.

Liu, Y., Goudreau, S., Oiamo, T., Rainham, D., Hatzopoulou, M., Chen, H., Davies, H., Tremblay, M., Johnson, J., Bockstael, A., others, 2020. Comparison of land use regression and random forests models on estimating noise levels in five Canadian cities. Environmental pollution 256, 113367.

Lundberg, S.M., Erion, G., Chen, H., DeGrave, A., Prutkin, J.M., Nair, B., Katz, R., Himmelfarb, J., Bansal, N., Lee, S.-I., 2020. From local explanations to global understanding with explainable AI for trees. Nat Mach Intell 2, 56–67.

Lundberg, S.M., Lee, S.-I., 2017. A unified approach to interpreting model predictions. Adv Neural Inf Process Syst 30.

Ma, X., Zou, B., Deng, J., Gao, J., Longley, I., Xiao, S., Guo, B., Wu, Y., Xu, T., Xu, X., others, 2024. A comprehensive review of the development of land use regression approaches for modeling spatiotemporal variations of ambient air pollution: A perspective from 2011 to 2023. Environ Int 108430.

Mann, S., Singh, G., 2022. Traffic noise monitoring and modelling—an overview. Environmental Science and Pollution Research 29, 55568–55579.

Meller, G., de Lourenço, W.M., de Melo, V.S.G., de Campos Grigoletti, G., 2023. Use of noise prediction models for road noise mapping in locations that do not have a standardized model: A short systematic review. Environ Monit Assess 195, 740.

Molnar, C., Casalicchio, G., Bischl, B., 2018. iml: An R package for interpretable machine learning. J Open Source Softw 3, 786.

Murphy, E., King, E.A., 2010. Strategic environmental noise mapping: Methodological issues concerning the implementation of the EU Environmental Noise Directive and their policy implications. Environ Int 36, 290–298.

Peris, E., Arguelles, M., 2023. Small-area analysis of social inequalities in exposure to environmental noise across four urban areas in England. Sustain Cities Soc 95, 104603.

R Core Team, 2024. R: A language and environment for statistical computing. Vienna, Austria: R Foundation for Statistical Computing.

Raess, M., Brentani, A., de Campos, B.L. de A., Flückiger, B., de Hoogh, K., Fink, G., Röösli, M., 2021. Land use regression modelling of community noise in São Paulo, Brazil. Environ Res 199, 111231.

Ragettli, M.S., Goudreau, S., Plante, C., Fournier, M., Hatzopoulou, M., Perron, S., Smargiassi, A., 2016. Statistical modeling of the spatial variability of environmental noise levels in Montreal, Canada, using noise measurements and land use characteristics. J Expo Sci Environ Epidemiol 26, 597–605.

Sekeroglu, B., Ever, Y.K., Dimililer, K., Al-Turjman, F., 2022. Comparative evaluation and comprehensive analysis of machine learning models for regression problems. Data Intell 4, 620–652.

Smith, M.G., Cordoza, M., Basner, M., 2022. Environmental noise and effects on sleep: an update to the WHO systematic review and meta-analysis. Environ Health Perspect 130, 76001.

Smola, A.J., Schölkopf, B., 2004. A tutorial on support vector regression. Stat Comput 14, 199–222.

Staab, J., Schady, A., Weigand, M., Lakes, T., Taubenböck, H., 2022. Predicting traffic noise using land-use regression—a scalable approach. J Expo Sci Environ Epidemiol 32, 232–243.

Staab, J., Stark, T., Wurm, M., Wolf, K., Dallavalle, M., Schady, A., Lakes, T., Taubenböck, H., 2023. Using CNNs on Sentinel-2 data for road traffic noise modelling, in: 2023 Joint Urban Remote Sensing Event (JURSE). pp. 1–4.

Štrumbelj, E., Kononenko, I., 2014. Explaining prediction models and individual predictions with feature contributions. Knowl Inf Syst 41, 647–665.

Wang, M., Beelen, R., Eeftens, M., Meliefste, K., Hoek, G., Brunekreef, B., 2012. Systematic evaluation of land use regression models for NO2. Environ Sci Technol 46, 4481–4489.

Welch, D., Shepherd, D., Dirks, K.N., Reddy, R., 2023. Health effects of transport noise. Transp Rev 1–21.

World Health Organization, 2018. Environmental noise guidelines for the European region. World Health Organization. Regional Office for Europe.

Wright, M.N., Ziegler, A., 2017. ranger: A fast implementation of random forests for high dimensional data in C++ and R. J Stat Softw 77, 1–17.

Xie, D., Liu, Y., Chen, J., 2011. Mapping urban environmental noise: a land use regression method. Environ Sci Technol 45, 7358–7364.

Xu, X., Ge, Y., Wang, W., Lei, X., Kan, H., Cai, J., 2022. Application of land use regression to map environmental noise in Shanghai, China. Environ Int 161, 107111.

Yeo, I.-K., Johnson, R.A., 2000. A new family of power transformations to improve normality or symmetry. Biometrika 87, 954–959.

Yin, X., Fallah-Shorshani, M., McConnell, R., Fruin, S., Franklin, M., 2020. Predicting fine spatial scale traffic noise using Mobile measurements and machine learning. Environ Sci Technol 54, 12860–12869.

Zhu, J.-J., Yang, M., Ren, Z.J., 2023. Machine learning in environmental research: common pitfalls and best practices. Environ Sci Technol 57, 17671–17689.

Zou, H., Hastie, T., 2005. Regularization and variable selection via the elastic net. J R Stat Soc Series B Stat Methodol 67, 301–320.